# Pure circular polarization electroluminescence at room temperature with spin-polarized light-emitting diodes

N. Nishizawa [a)], K. Nishibayashi, and H. Munekata [b)]

Institute of Innovative Research [c)], Tokyo Institute of Technology

4259-J3-15 Nagatsuta, Midori-ku, Yokohama 226-8503, Japan

## ABSTRACT

We report the room-temperature electroluminescence (EL) with nearly pure circular polarization (CP) from GaAs-based spin-polarized light-emitting diodes (spin-LEDs). External magnetic fields are not used during device operation. There are two small schemes in the tested spin-LEDs: firstly, the stripe-laser-like structure that helps intensifying the EL light at the cleaved side walls below the spin injector Fe slab, and secondly, the crystalline $AlO_x$ spin tunnel barrier that ensures electrically stable device operation. The purity of CP is depressively low in the low current density ($J$) region, whereas it increases steeply and reaches close to the pure CP when $J > 100$ A/cm$^2$. There, either right- or left-handed CP component is significantly suppressed depending on the direction of magnetization of the spin injector. Spin-dependent re-absorption, spin-induced birefringence and optical spin-axis conversion are suggested to account for the observed experimental results.

[a)] Electronic mail: nishizawa.n.ab@m.titech.ac.jp and [b)] mail: munekata.h.aa@m.titech.ac.jp. c) Formally, Imaging Science and Engineering Laboratory.

## Significance

Most of the experiments on the spin manipulation in semiconductors, the principal materials in modern electronic and photonic devices, were carried out at cryogenic temperatures and high magnetic fields, because thermal energy tends to randomize spin information in the semiconductor that is non-magnetic. Here, we report very surprising experimental results of pure circular polarization electroluminescence at room temperature with no external magnetic fields. They are obtained by electrically injecting moderately high density of spins into semiconductor double heterostructures, the structures that were invented in connection with semiconductor lasers half-century ago. The results suggest the appearance of some spin-dependent non-linear processes that lead to recovering and even enhancing the injected, initial spin information in semiconductors.

¥body

As well represented by the giant magneto-resistance (GMR), tunneling magneto-resistance (TMR), and spin-transfer-torque magnetic random access memory (STT-MRAM), spintronics research based on spin transport in magnetic metals has been contributing significantly in the progress of electronics through the advancement in recording bit density and low-power memory retention [1, 2]. Proposal of the spin current modulation with an electric field [3] and invention of diluted magnetic III-V semiconductors [4] have opened the opportunity of introducing spin degree of freedom in semiconductor technology [5]. After those works, light-induced magnetism [6], electric-field-controlled magnetism [7], spin qubits in semiconductors [8, 9], spin-polarized light emitting diodes (spin-LED) [10, 11], and spin-MOSFET [12] were either demonstrated or proposed, which have caused not-a-small impact on the metal-based spintronics and applied physics. However, works that assure the room-temperature (RT) operation of those semiconductor-based devices have not been accomplished to date.

Concerning the spin-LED, the highest CP value, $P_{CP} \equiv \{I(\sigma^+)-I(\sigma^-)\}/\{I(\sigma^+)+I(\sigma^-)\}$with $I(\sigma^+)$ and $I(\sigma^-)$ the intensity of right- and left-handed EL component, respectively, was $P_{CP} \approx$ 0.3 ~ 0.35 at RT in a 0.8-T magnetic flux ($B$), which was achieved in the context of studying the spin-filtering effect of the MgO tunnel barrier [13, 14]. Most of the past works regarding spin-LED were carried out under the vertical arrangement with low $J$ ranging 0.1 - 1 A/cm$^2$ and forcing spins aligned vertically by applying out-of-plane external magnetic fields [15].

With vertical-cavity surface-emitting laser structure incorporating the quantum wells (QWs) [16-18] or thick active layer [19] together with a means of a vertical optical resonator, pure-CP lasing was demonstrated by the optical pumping up to RT [16-19]. On the other hand, the CP lasing achieved by the electrical pumping have been $P_{CP} \sim 0.23$ at 50 K with a direct current of $J \sim 2.8$ kA/cm$^2$ and $B \sim 2.0$ T in InGaAs-based QWs [20], $P_{CP} \sim 0.55$ at 230 K with a pulsed-current of presumably $J \sim 18$ kA/cm$^2$ and $B \sim 2.0$ T in InGaAs-based quantum-dots [21], and $P_{CP} \sim 0.28$ at RT with pump energy of 87 μJ and $B \sim 0.35$ T in GaN nanorods filled with $Fe_3O_4$ nanoparticles [22].

Here, we report the unforeseen appearance of nearly pure CP electroluminescence, $P_{CP} =$ 0.95 at RT from the edge-emission-type spin-LED structure that was less investigated in the past. Our results suggest the appearance of some non-linear effect in the regime of moderately high current density ($J > 100$ A/cm$^2$), and lead us to the opportunity of studying the RT operation of, at least, semiconductor-based spin-photonic devices.

Shown in Fig. 1*A* is a schematic cross section, the cleaved GaAs (110) side wall, of the tested LED chips. They consist of a polycrystalline Fe in-plane spin injector, a crystalline $\gamma$-like $AlO_x$ tunnel barrier [23], and an epitaxial AlGaAs/GaAs/AlGaAs double heterostructures (DHs) [24]. A magnetization vector of the spin injector is controlled either along parallel or anti-parallel to the GaAs [110] axis by the technical magnetization. Spin polarization of electrons, $P_e = (n^+ - n^-)/(n^+ + n^-)$ with $n$ the electron density at the Fermi level, is assumed $P_e$

≈ 0.4 in Fe [25]. At the stage of initial injection, 70 % of electron spins point toward the GaAs [110] axis that is superposition of the two primary crystal axes, [100] and [010], whereas remaining 30 % have the opposite, $[\bar{1}10]$ spin axis. Spin/charge transport takes place vertically toward the [001] axis.

The DH wafer was designed by the authors in view of (i) not severely reducing the spin polarization during carrier transport across the upper *n*-AlGaAs clad layer and (ii) avoiding large optical loss due to the top spin injector metal. Taking account of these two contradictory requirements, the thickness and alloy composition of the $n$-$Al_xGa_{1-x}As$ layer was chosen to be $L$ = 500 nm and $x$ = 0.3, respectively. With this layer, 60 % of the injected $x$-axis spins are supposed to preserve their spin axis at the *p*-GaAs active layer [26, 27], whereas around 4 % of the EL energy is absorbed in the Fe injector, assuming the extinction coefficient $\kappa = 3.89$ at the wavelength 909 nm [28] and the confinement efficiency of DH $\Gamma = 0.91$ [29]. A thick GaAs active layer, in which heavy- and light-hole bands are degenerated, is used in order to accommodate spins of in-plane axes in the active layer. Moreover, the active layer is intentionally doped *p*-type in order to reduce the radiative recombination lifetime [30] and suppress the contribution of non-radiative recombination. Note that initially injected [110] (or $[\bar{1}10]$) spins are dispersed into the spins of other orthogonal orientations ($[\bar{1}10]$ and [001]) as a result of spin relaxation. Radiative recombination of these spins is observed as the linearly polarized emission: namely, the helicity-independent background, in the CP resolved EL detection system whose optical axis is along the [110] axis (Fig. 2*A*).

The DH part was grown on a *p*-GaAs (001) wafer using a metal-organic vapor phase epitaxy reactor at the Optowell Co. Ltd. in order to ensure high optical quality that the radiative recombination dominates at RT. A 1-nm thick crystalline $\gamma$-like $AlO_x$ tunnel barrier was then grown by the authors using molecular beam epitaxy chamber [23]. The density of interface states at the $AlO_x$/GaAs interface has been found $D_{it} \approx 3 \times 10^{11}$ $cm^{-2} \cdot eV^{-1}$ [31], which is far less than that at the amorphous $AlO_x$/GaAs interface. This was followed by the fabrication of 100-nm thick, 40-μm wide Au (20 nm) / Ti (5 nm) / Fe(100 nm) spin injector stripes on top of the tunnel barrier using a separate e-beam evaporator and standard photolithography. Finally, the wafer was thermally annealed at 230 °C for 60 min in the nitrogen gas atmosphere, and then cleaved into 1.1 × 2.0 mm rectangle chips. The resultant aspect ratio of the spin injector is 1:25. The long side of the injector, the easy axis, is parallel to the short side of the chip which is the GaAs [110] axis. They were mounted on a copper block for the electroluminescence (EL) experiments.

EL spectra obtained from the control experiment using the chip without a Fe layer is shown in Fig. 1*B*, together with photoluminescence (PL) spectrum obtained by the surface excitation of the DH wafer. The peak energy at 1.43 eV in the PL spectrum indicates the domination of near-band-edge emission, whereas, in the EL spectrum, the near-band-edge emission is weak and the peak appears at the photon energy that is around 100 meV less than that in the PL spectrum. That is to say, reflecting a relatively long, lateral optical path, the near-band-edge emission is significantly re-absorbed due to a large absorption coefficient ($\alpha \geq 10^4$ $cm^{-1}$ ), whereas the low-energy emission ($h\nu \leq 1.38$ eV), which is attributed to the transition from the conduction band to the valence-band tail states caused by the mixing with acceptor states [33], is less re-absorbed owing to a small absorption coefficient ($\alpha < 10^2$ $cm^{-1}$ ). The right- and left-handed EL spectra are identical to each other, indicating $P_{CP} = 0$.

Prior to EL measurements, the Fe spin injector was magnetized along the long side of the stripe ([110]) by an external magnetic field of $H = 5$ kOe. The EL emitted from the cleaved (110) side walls was transmitted through a quarter-wave plate (QWP) and a linear polarizer (LP), and was collected into a multi-channel spectrometer (MCS) that was placed 30 cm away from the cleaved edge (Fig. 2*A*). Setting the optical axis of QWP either at 0° or 90°, while fixing the optical axis of LP at 45°, either right- ($\sigma^+$) or left- ($\sigma^-$) handed EL component was selected, respectively. This method excludes confusion with the linearly polarized EL that composed of co-radiation of transverse-magnetic and transverse-electronic modes. In fact, as shown in Fig. 2*B*, the intensity relation between the right-handed (QWP at 0°) and the left-handed (QWP at 90°) components is reversed when the direction of remnant magnetization is reversed: e.g., for the emission band around $h\nu = 1.35$ eV, the $P_{CP}$ values ($P_{CP} \equiv \{I(\sigma^+)-I(\sigma^-)\}/\{I(\sigma^+)+I(\sigma^-)\}$) are 0.83 and −0.80 for the magnetization pointing toward (+***M***) or against (−***M***) the QWP, respectively. The $I$ value hereafter is the emission intensity integrated over the photon energies within the full-width at half maximum of the EL emission band.

Electroluminescence occurs when the *p*-GaAs substrate side is positively biased above around 1 V. Increasing the bias results in an increase of a current as well as the emission intensity especially from the side-wall area underneath the spin injector stripe. A representative $J$ (current density) – $V$ (voltage) curve from the chip A is shown in Fig. 3*A* and 3*B* in linear and semi-logarithmic scales, respectively. Here, $J$ is estimated using the spin injector area, 40 μm × 1.1 mm, assuming the short current spreading length (~ 2 μm) in the lateral direction [34]. The slope of the $J - V$ curve, $eV / nkT$, at the relatively high bias region ($V \geq 1.3$ V) shows the diode factor $n \sim 2$, indicating the recombination-dominated carrier transport around the *p*-*n* junction region [35].

Fig. 3*C* shows EL spectra obtained at three different $J$ values, 22, 55, and 110 A/cm$^2$. Two relatively broad emission bands are present: the one with the peak around 1.42 eV (band A) is due to the band-to-band emission, and the other peaking around 1.36 eV (band B) is attributed to the emission via the valence band tails that propagates laterally through the GaAs active layer [36]. Comparing these spectra with that of the control experiment (Fig. 1*B*), the appearance of the band A and the blue shift of the band B are noticeable. These facts can be understood in terms of the difference in the extinction coefficient between Fe and Au layers; $\kappa = 3.89$ and 6.06 at the wavelength 909 nm for Fe and Au, respectively [28, 32]. The band A

(B) shifts toward higher (lower) energies when $J$ is increased; the peak photon energies of the band A is 1.406, 1.4065, and 1.414 eV, whereas those of the band B 1.365, 1.360, and 1.351 eV, at $J = 22$, 55 and 110 A/cm$^2$, respectively. Utilizing the observed blue and red shifts at $J =$ 110 A/cm$^2$, electron density and the temperature in the $p$-GaAs active layer are estimated to be around $6\times10^{17}$ cm$^{-3}$ and around 319 K, respectively [37]. Note that this electron density is around one fifth (1/5) of the threshold density for CP lasing that has been estimated by the optical pumping experiments using the VCSEL incorporating a bulk, undoped GaAs active layer ($d = 485$ nm) [19].

It is clearly noticeable that the difference in intensity between the $\sigma^+$ and $\sigma^-$ EL components becomes larger with increasing the current density $J$. The spectral shape, however, is nearly identical between the two components (Inset in Fig. 3*C*). As to the CP value, which is defined previously by the form $P_{CP} \equiv \{I(\sigma^+)-I(\sigma^-)\}/\{I(\sigma^+)+I(\sigma^-)\}$, $P_{CP} = 0.14$, 0.42, and 0.95 for the band A, whereas $P_{CP} = 0.11$, 0.44, and 0.94 for the band B, at $J = 22$, 55, and 110 A/cm$^2$, respectively. Bias dependence on the $P_{CP}$ value for the band B is summarized in Fig. 3*D*. The observed increase in the $P_{CP}$ value at relatively low bias region ($V < 6$ V) verifies the theoretically proposed spin-injection-impeding effect that is supposed to appear in the depletion region in semiconductor junctions [38]. Spectral dependent $P_{CP}$ data are presented in the supporting information section including the data from other chips. Those results suggest that there is no essential difference between majority and minority spins in terms of the electronic states associated with the electron-hole recombination. In terms of the relation in emission intensity between the bands A and B, the ratio $I_B/I_A$ gradually increases with increasing $J$ irrespective of the $\sigma^+$ and $\sigma^-$ EL components: $I_B/I_A = 1.1$ and 2.2 at $J = 20$ and 100 A/cm$^2$, respectively. Namely, the band B becomes dominant at high $J$ region.

Reduction in the intensity of the $\sigma^-$ (minority) component $I(\sigma^-)$ with increasing $J$ are the common trend for the tested chips that have survived in the region $J \geq 50$ A/cm$^2$. Representative intensity data obtained from three different chips are shown in Fig. 4*A* for the emission band B. The value of $I(\sigma^-)$ tends to saturate in the region $J \approx 40$ - 80 A/cm$^2$ beyond which it decreases, whereas the $\sigma^+$ (majority) component $I(\sigma^+)$ increases nearly linearly throughout the entire $J$ region within the limit of the present work. Consequently, nearly pure CP is realized when $J$ reaches around 100 A/cm$^2$ or higher (Fig. 4*B*). Moderate narrowing of the EL band B has also been observed for these chips in the region $J > 50$ A/cm$^2$. (Fig. 4*C*).

At the point of writing this report, EL intensity and CP value both tend to degrade in around half day of operation when $J > 100$ A/cm$^2$. The origin of the observed degradation has not been elucidated by metallurgical evaluations with cross sectional transmission electron microscope and secondary ion mass spectrometry. We suppose that it is originated from some slow degradation of a crystalline γ-$AlO_x$ tunnel barrier, which should be improved in the future study. At least it is clear that, without the $AlO_x$ barrier, chips only show poor EL performance as disclosed in the supporting information section.

Figure 5 shows the horizontal line profiles of the $P_{CP}$ value and integrated EL intensity; the point $y = 0$ represents the position of the cleaved side wall right under the Fe strip center. Measurements were carried out by laterally moving the 0.1-mm wide, 10-mm long, vertical optical slit that was placed 0.1 mm away from the cleaved side wall. At $J = 75$ A/cm$^2$, the $P_{CP}$ value, which maximizes at $y = 0$ with $P_{CP} = 0.18$, drops abruptly and becomes nearly zero at $y \sim 0.1$ mm, whereas the EL intensity profile extends out to larger $y$ values. At $J = 125$ A/cm$^2$, the $P_{CP}$ value increases significantly at $y = 0$ and decreases nearly zero at $y \sim 0.2$ mm. Interestingly, the intensity profile is narrowed within $y \sim 0.2$ mm, suggesting the concentration of EL emission energy in the region under the stripe electrode. We infer that this phenomenon is relevant to the observed significant increase in the $P_{CP}$ value.

Suppression of the spin-injection-impeding effect [38] by a large forward bias will enhance the electron spin polarization in the active region up to $P_e \sim 0.24$, and will result in the EL with the $P_{CP}$ values ranging between 0.12 and 0.24. The lower and upper bounds of $P_{CP}$ correspond to the case in which the spin-dependent optical selection rule [39] (Fig. 6*B*) fully applies and relaxes, respectively. On the other hand, as discussed previously, the density of non-equilibrium electrons in the active layer is estimated to be in the range of middle $10^{17}$ cm$^{-3}$ at $J \sim 100$ A/cm$^2$, which is around one fifth (1/5) of the threshold value for the CP lasing in an undoped GaAs layer [19]. The threshold value in our chips is presumably even larger, referring to the fact that the lifetime of radiative recombination in $p$-GaAs ($5 \times 10^{-9}$ sec at $p \sim 10^{18}$ cm$^{-3}$) is two orders of magnitude shorter than that in lightly-doped $p$-GaAs ($> 1 \times 10^{-6}$ sec with $p < 5 \times 10^{15}$ cm$^{-3}$) [30]. These comparisons lead us to the inference that the mechanism of the observed suppression of the minority CP EL component at $J > 100$ A/cm$^2$ is not originated from effects associated with the spin-injection-impeding effect [38] or the spin-polarized semiconductor lasers [40, 41]; in the latter, nearly pure circular polarization appears when majority spins enter the regime of stimulated emission whereas minority spins stay in the spontaneous emission regime.

In our experiments, the variation in the EL intensity as a function of $J$ is higher for the emission band B than that for the emission band A. We examine this fact in detail on the basis of microscopic process as depicted in Fig. 6*A*. A part of electrons injected in the conduction band of the active layer undergo the transitions to the empty states in the valence band that includes the band-tail states, as represented by the process A in the figure. Here, the Fermi level is positioned around the valence band edge. During this process, electrons in the valence band are expelled from the active layer as the consequence of ultrafast dielectric relaxation; its time constant is estimated to be $\tau_D = \varepsilon \cdot \varepsilon_0 / \sigma \sim 60$ fs in which $\varepsilon_0$ is the dielectric constant of the vacuum, $\varepsilon$ the relative dielectric constant ($\varepsilon = 13.2$), and $\sigma$ the conductivity ($\sigma \sim 20$ $\Omega^{-1} \cdot$cm$^{-1}$). This process, as denoted by the process B in the figure, is alternatively called hole injection in the valence band. In parallel with this process, electrons that are captured by the tail states relax toward the empty states near the Fermi level $E_F$ with the time constant in the range of sub-ps [42] (process C). In the meantime, a part of photons generated by the process A are re-absorbed by the electrons near $E_F$ (the process D) within the residual time of photons that is around 10 ps for the lateral optical pass length of 1 mm. The rate of re-absorption is higher for photons whose energy is comparable to or higher than the intrinsic band gap ($E_g$ = 1.44 eV, $\alpha > 5 \times 10^3$ cm$^{-1}$) than those of $h\nu < 1.44$ eV ($\alpha < 10^2$ cm$^{-1}$), which results in the $J$ dependent $I_B/I_A$. The dynamics of the entire EL process is completed by adding the non-radiative recombination process E. Rate equations are shown in the supporting information section.

We next introduce spins on the carrier dynamics. We then notice that spins expelled by the dielectric relaxation from the active layer still preserve their own spin polarization, because the lifetime of spins in the valence band is longer, being $\tau_S \sim 900$ fs [43, 44], than the $\tau_D$ value. This naturally means that carriers near the Fermi level are spin polarized. The magnitude of spin polarization ($P_h$) is negligibly small when the number of non-equilibrium electrons ($N^*$) reaching the empty states in the valence band via the process A is small, as is the case for the low $J$ region. $P_h$ becomes noticeably high when $N^*$ reaches the value that is comparable to the background hole concentration, $10^{18}$ cm$^{-3}$ in the present case, as is the case for the high $J$ region. Note that the polarization of $P_h$ is opposite from that of $P_e$, since the same amount of spins entering the conduction band are expelled from the valence band. Turning eyes on the re-absorption, the $\sigma^-$ EL component that consists of minority-spin photons is more absorbed than the $\sigma^+$ EL component, reflecting the reversed spin population of carriers in the valence band. Thus, the observed reduction in the $\sigma^-$ EL component at high $J$ region can be explained qualitatively in terms of the dielectric relaxation followed by the re-absorption.

For an undoped active layer, the contribution of non-radiative recombination is presumably still significant at $J \sim 100$ A/cm$^2$ since the lifetime of radiative recombination is around two orders of magnitude longer [30]. We suppose that the effect associated with the reversed spin population is not observed unless $J$ reaches near the stimulated emission condition. It is likely that $p$-type doping in the active layer has led us to the observation of pure CP at $J \sim 100$ A/cm$^2$. This inference also calls interesting questions concerning the blueprint of spin-LED; *e.g.*, how much is the appropriate $p$-type doping level and how much is the lowest bound of the $P_e$ value for the electrons injected in the active layer? In order to find quantitative answers, it is desired to carry out the model calculation based on couples of spin-polarized rate equations that describe dynamics of electrons in the conduction band, in the tail states, and around the Fermi level, as well as photons, incorporating the spin-dependent optical selection rule, in addition with further experimental investigations.

Let us next infer the possible non-linear effects in the optical process. Referring the optical selection rule mentioned previously (Fig. 6*B*), radiative recombination results in the elliptic polarization (EP), the superposition of $\sigma^+$ and $\sigma^-$ EL components. Since GaAs does not show birefringence under the normal condition, the relative phase difference between the two orthogonal light waves having the electric fields $E_y$ and $E_z$ in the rectangular coordinate system shown in Fig. 1,

$$\begin{pmatrix} E_y \\ E_z \end{pmatrix} = \begin{pmatrix} A_1 \exp\{i(kx - \omega t + \phi_1)\} \\ A_2 \exp\{i(kx - \omega t + \phi_2)\} \end{pmatrix}$$

is unchanged during the light propagation. If one assumes the inducement of birefringence $\begin{pmatrix} 1 & +\delta \\ -\delta & 1 \end{pmatrix}$ due to spin-polarized carriers, the relative phase delay occurs between the $E_y$ and $E_z$ waves when EL light propagates along the GaAs active layer under the Fe slab electrode in which spin-polarized carriers with the density of middle $10^{17}$ – $10^{18}$ $cm^{-3}$ are injected. Consequently, the ratio between $\sigma^+$ and $\sigma^-$ components of the light coming outside the active layer along the $x$ axis differs from the original EP condition. Referring the experimental results shown in Figs. 3 and 4, the relative phase delay between $E_y$ and $E_z$ is supposed to be spin dependent: the ± sign of $\delta$ in the off diagonal term in the dielectric tensor is swapped by altering the spin axis, whereas its magnitude varies with injected spin density. This scenario explains well the almost identical EL spectra of the two different helicity (inset Fig. 3*C* and Fig. 1S in supporting information section). On the other hand, this scenario requires the de-randomization of optical path length from the point of electron-hole recombination radiation to the cleaved side wall. An increase in the effective optical path length would occur when stimulated emission takes place in part in the active region. The observed moderate spectral and spatial narrowing at $J > 100$ A/cm$^2$ (Fig. 4*C* and Fig. 5) may be a faint indication of the precursory stage to the stimulated emission. The difference in refractive indices is estimated to be $\Delta n = 2 \times 10^{-4}$ assuming the phase delay of $\pi/2$ in the optical path of 1 mm, which is rather small in view of the birefringence materials standard [45].

It is worth addressing another interesting view that is associated with the radiative recombination of the $[1\bar{1}0]$ and [001] electron spins with valence-band holes. Recall that the light emission due to such recombination is detected as the linearly polarized background of equal intensity in the present measurement setup (Fig. 2*A*). If some sort of non-linear process converts the linearly polarized background into CP emission at moderately high *J*, the observed decrease (increase) in minority- (majority-) CP component may occur (Fig. 4*A*). This scenario would rather be likely to take place in the bulk-type active medium than in the quantum-well-type medium in which spin axis is quantized in one particular direction (the GaAs [001] axis).

Although the present work serves the solid footstep toward the development of semiconductor-based, spin-photonic devices, it is just the beginning of our journey. Investigation at higher *J* region is extremely interesting since, not only in view of elucidating the mechanism of non-linear effects, but also to clarify whether stimulated emission with circular polarization is possible in the lateral-waveguide-type geometry in which stimulated light emission with linear polarization dominates. Exploitation towards both shorter and longer wavelengths with different materials combinations, together with electrical CP switching [46], is another important direction in view of investigating the usefulness of spin-LEDs in the existing and new applications, e.g., the chiral resolution in synthetic chemistry [47], diagnosis of cancerous tissues [48],circularly polarized ellipsometry [49], the optically enhanced nuclei imaging [50], LP-polarization-insensitive three-dimensional display [51], quantum eraser technique [52], optical secure communications [53], and beyond.

**Acknowledgements**

Authors are grateful to M. Aoyama for his technical assistance in various experiments, and F. Minami, F. Koyama, T. Miyamoto, and S. Arai for various discussions concerning light emitting devices. We acknowledge financially supports from Advanced Photon Science Alliance Project from MEXT and Grant-in-Aid for Scientific Research (No. 22226002) from JSPS.

## Figure legends

**Fig. 1** (*A*) A schematic cross section of spin-LEDs. (*B*) EL and PL spectra obtained from the control chip consisting of Au/Ti/$\gamma$-like $AlO_x$/DHs. Solid lines represent EL spectra with three different current densities, $J = 100$ ($\sigma^+$, red and $\sigma^-$ black), 75 (green) and 50 (blue) A/cm$^2$. The range of applied voltage is between 2 to 4 V. A black, dotted line shows PL spectrum obtained by the surface excitation ($h\nu = 1.58$ eV, $\lambda = 785$ nm, 40 mW). The extinction coefficient of a top Au electrode is $\kappa = 6.06$ at 1.36 eV ($\lambda = 909$ nm) [32].

Fig. 2 (*A*) A schematic experimental setup for EL measurements, showing, from upper left to lower right, a wire-bonded rectangle spin-LED chip on a copper block, a quarter-wave ($\lambda/4$) plate (QWP), a linear polarizer (LP), and a multi-channel spectrometer (MCS). A pair of lens, one in between the chip and QWP and another in between LP and MCS, is omitted for graphical clarity. Orange waves represent EL from the chip with right- ($\sigma^+$, a red circle) and left- ($\sigma^-$, a blue circle) handed EL components. Straight orange arrows accompanied by both-headed arrows represent light waves converted into linear polarization by QWP. Thin dotted, blue arrows on polarizers represent optical axes. Pictures of chip A with EL from the cleaved edge are shown in the inset; current density $J = 22$ (left) and 110 A/cm$^2$ (right). (*B*) A couple of helicity-specific EL spectra obtained when direction of remnant magnetization

points toward QWP ($+M$, upper panel) and against QWP ($-M$, lower panel).

**Fig. 3** $J - V$ curves of the chip A in (*A*) linear scale and (*B*) semi-logarithmic scale, together with (*D*) a plot CP value ($P_{CP}$) against $J$. Somewhat larger bias voltage compared to that in the control sample suggests the formation of interface resistance in the Au/Ti/Fe/γ-AlOx electrode. (*C*) Helicity-specific EL spectra obtained at RT from a cleaved side wall of the chip at three different current densities, $J$ = 22 (green), 55 (blue), and 110 (red) A/cm$^2$, respectively. Solid and dotted lines show right- ($\sigma^+$) and left- ($\sigma^-$) handed components, respectively. Three vertical arrows and dotted lines in (*A*), (*B*) and (*D*) represent the $J$ values at which EL spectra are measured. The EL spectra measured at $J$ = 110 A/cm$^2$ are re-plotted in semi-logarithmic scale in the inset (*C*).

**Fig. 4** Plots of (*A*) the integrated intensities of right-handed ($\sigma^+$, closed symbols) and left-handed ($\sigma^-$, open symbols) components, (*B*) $P_{CP}$ values, and (*C*) spectral width of the band B represented by full-width at half maximum (FWHM) values, as a function of current density $J$ for three different spin-LED chips.

**Fig. 5** Horizontal line profiles of the $P_{CP}$ value and integrated EL intensity obtained from the chip C at $J$ = (*A*) 75 and (*B*) 125 A/cm$^2$. The point $y$ = 0 represents the position of a cleaved

side right under the center of the Fe strip electrode, as shown in the inset. Measurements were carried out by laterally moving the 0.1-mm wide, 10-mm long, vertical optical slit that was placed 0.1 mm away from the cleaved edge.

**Fig. 6** (*A*) A schematic illustration of transport - re-absorption scenario. The labels "Rad. rec." and "Non-rad. rec." represent radiative and non-radiative recombination processes, respectively. A red curve denotes the Fermi distribution function around the Fermi level ($E_F$) that is shown by a dashed-dotted line. Hatched areas depict the states occupied by electrons (red) and holes (blue) above and below the Fermi level, respectively. Inside the area surrounded by a dotted line represents an active layer, whereas the outside denotes cladding layers (upper and lower spaces) and a free space (right and left spaces).

(*B*) A schematic illustration that shows non-linear effects in the optical process; electrical spin injection from a Fe spin injector into a three-dimensional (3D) GaAs active layer through a 3D, *n*-AlGaAs clad layer (upper panel), radiative recombination of [110], $[1\bar{1}0]$ and [001] spins in C.B. with degenerated heavy- and light-holes in V.B. (lower left panel), and conversion from elliptic polarization into pure circular polarization through hypothetical non-linear optical process (lower right panel).

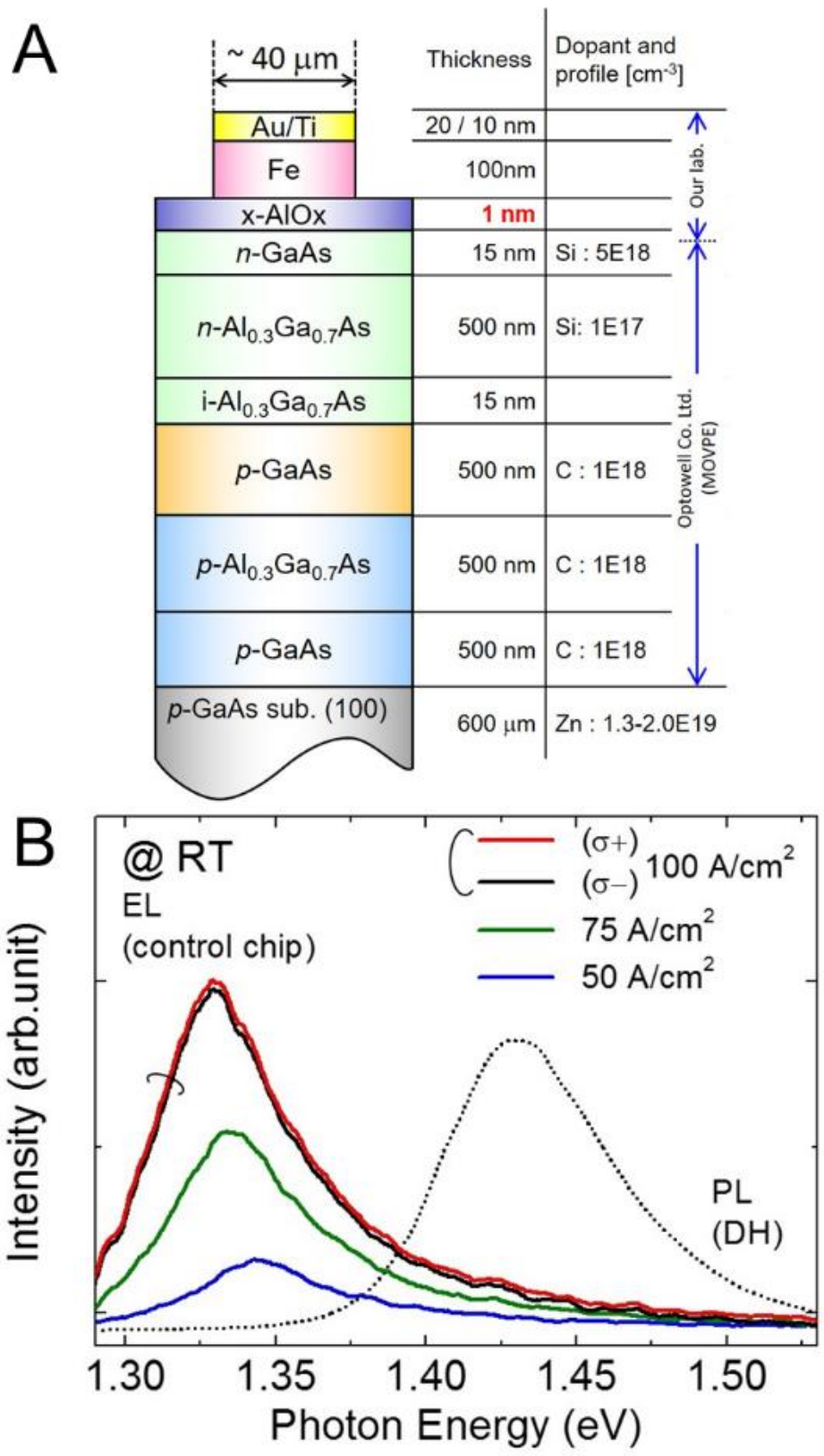

Fig. 1 N. Nishizawa *et al.*

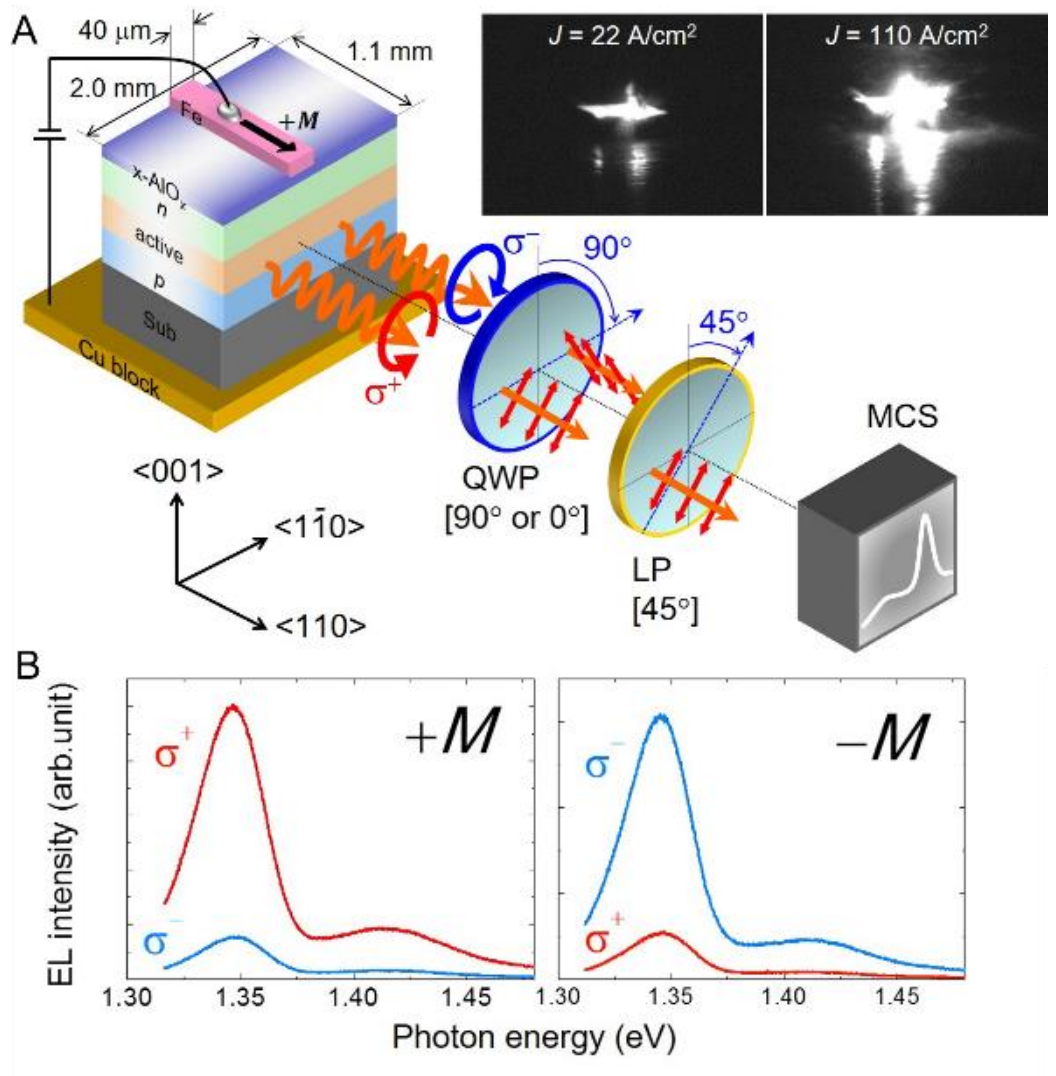

Fig. 2 N. Nishizawa *et al.*

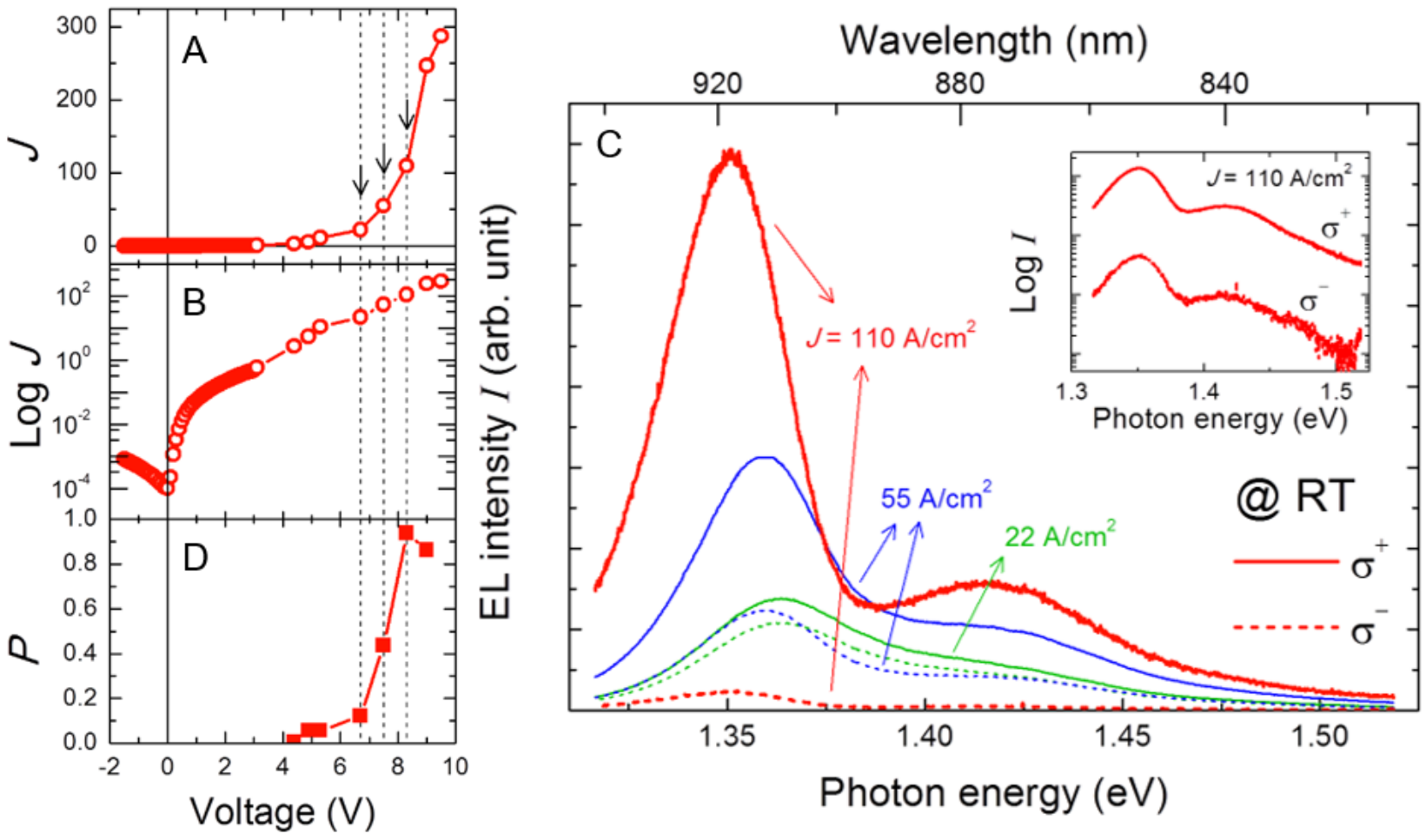

Fig. 3 N. Nishizawa *et al.*

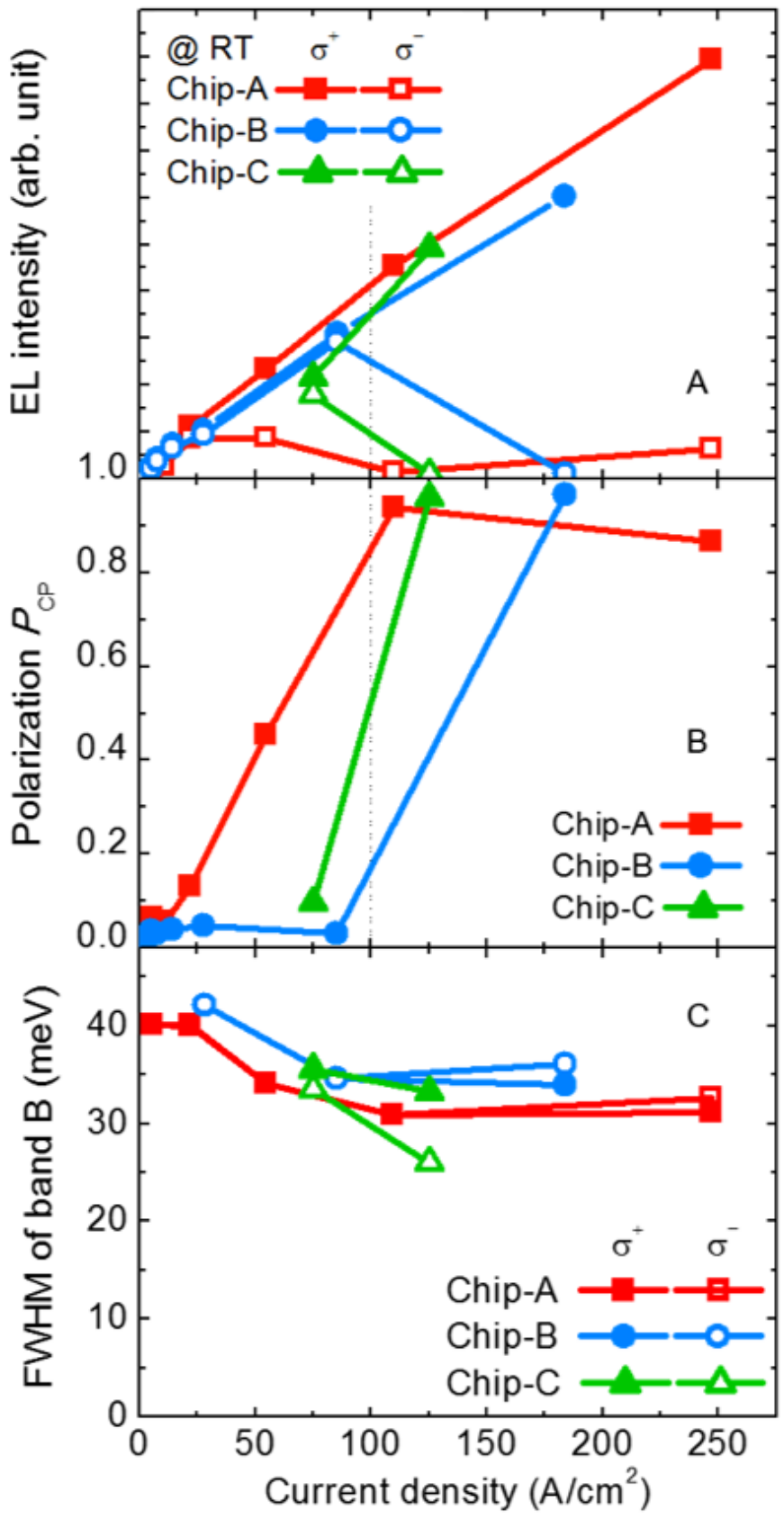

Fig. 4 N. Nishizawa *et al.*

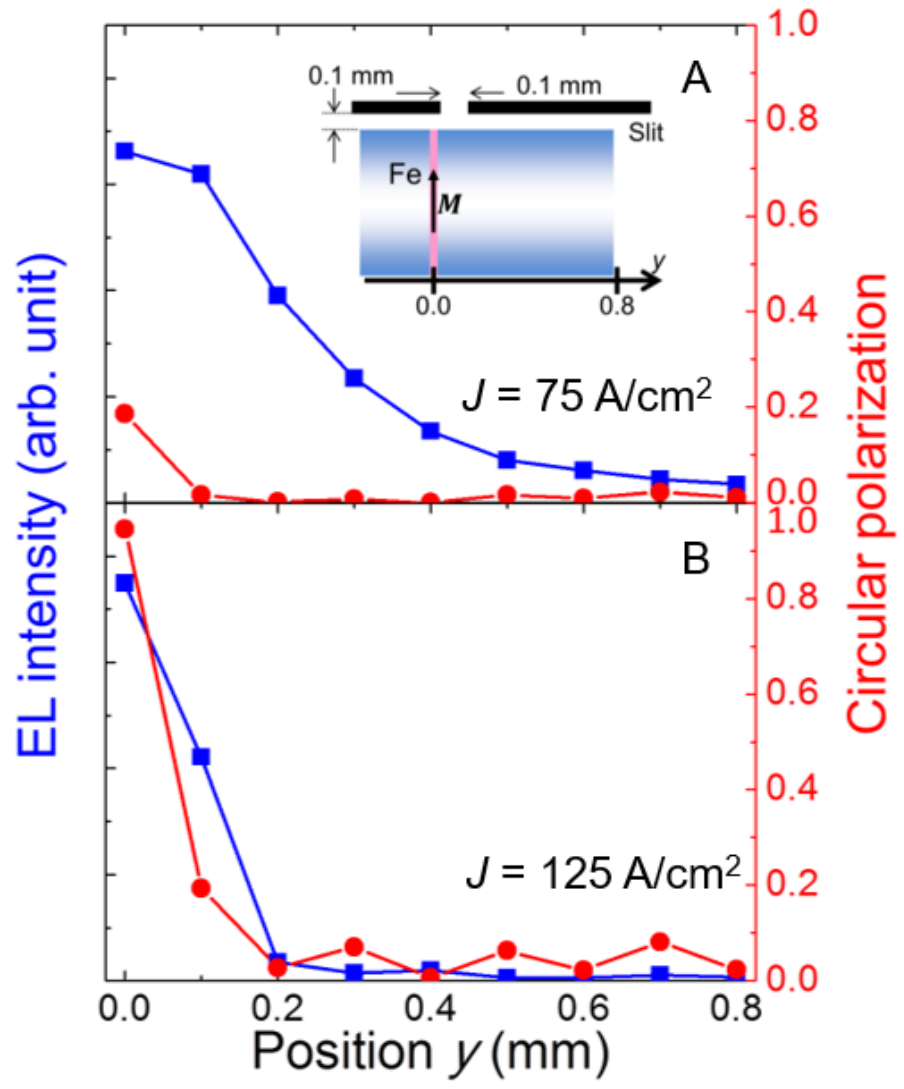

Fig. 5 N. Nishizawa *et al.*

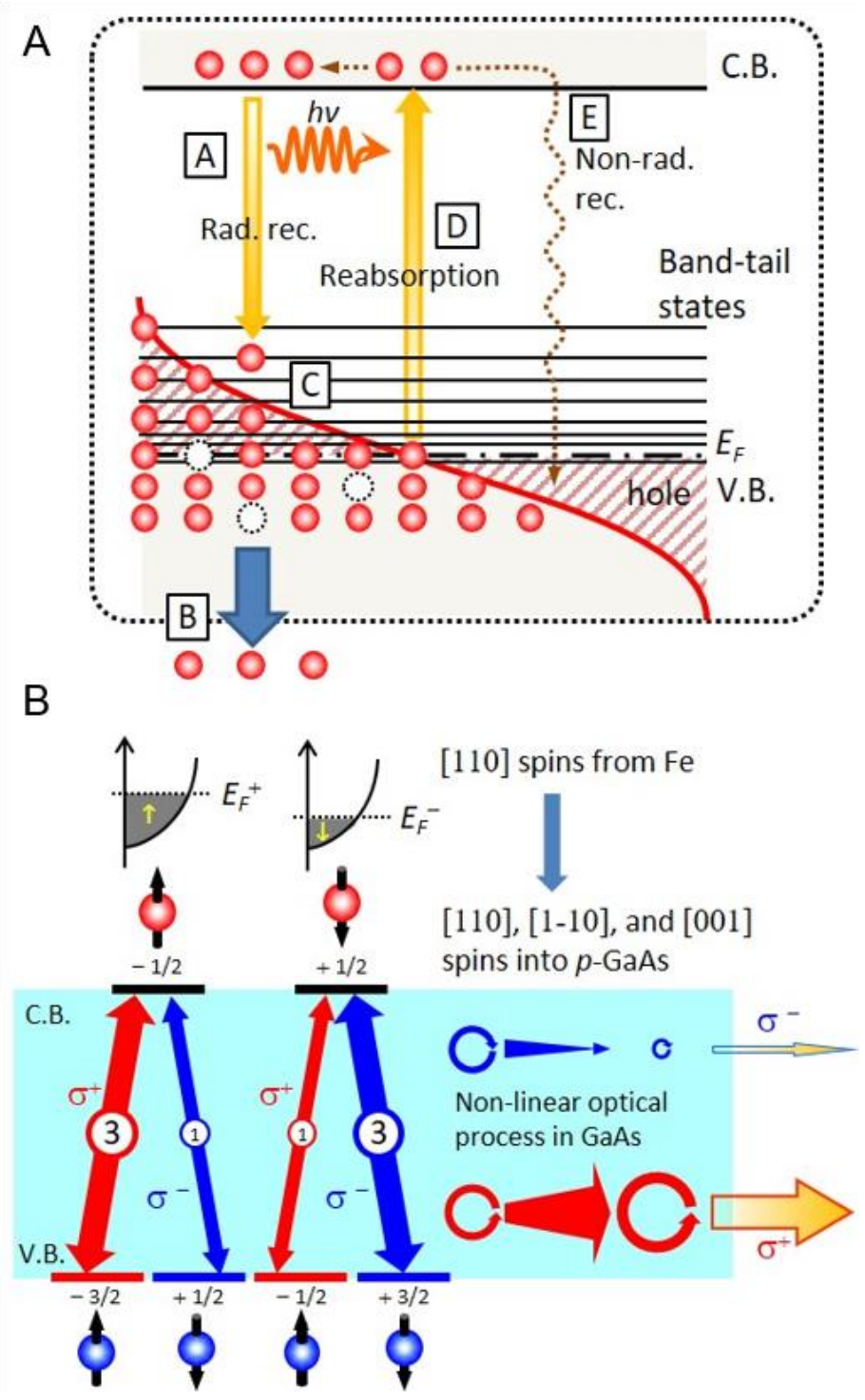

Fig. 6 N. Nishizawa *et al.*